\documentclass[aps,prl,twocolumn,showpacs,superscriptaddress]{revtex4}
\usepackage{epsfig}
\usepackage{graphicx}
\begin{document}
\topmargin-1.0cm

\title {
The role of triplet states in the emission mechanism of polymer
light-emitting diodes}

\author {M. Arif}\affiliation {Department of Physics and Astronomy, University of Missouri, Columbia, Missouri 65211, USA}\author {S. Mukhopadhyay}\author {S. Ramasesha}
\affiliation {Solid State and Structural Chemistry Unit, Indian Institute of Science, Bangalore 560012, India} \author{S.
Guha}\email[Corresponding author E-mail:]{guhas@missouri.edu} \affiliation {Department of Physics and Astronomy, University of Missouri,
Columbia, Missouri 65211, USA}

\date{\today}
\begin{abstract}
The blue emission of polyfluorene (PF) based light-emitting diodes
(LEDs) is known to degrade due to a low energy green emission, which
hitherto has been attributed to oxidative defects. By studying the
electroluminescence from ethyl-hexyl substituted PF LEDs in the
presence of oxygen and in an inert atmosphere, and by using trace
quantities of paramagnetic impurities (PM) in the polymer, we show
that the triplet states play a major role in the low energy emission
mechanism. Our time-dependent many-body studies show that there is a
large cross-section for the triplet formation in the electron-hole
recombination process in presence of PM, and intersystem crossing
from excited singlet to triplet states.

\end{abstract}

 \maketitle
%******************************************************************************
%\section{} \label{}
Defects in organic semiconductors impede charge transport and
emission mechanisms in organic light-emitting diodes (LEDs).
Emission energies lower than the $\pi$-$\pi$* gap in conjugated
polymers (CP) have been attributed to chemical and structural
defects, aggregates and interchain interactions. Polyfluorenes (PF)
have emerged as an especially attractive CP due to their strong blue
emission, high charge carrier mobility, and thus great prospects for
device application \cite{neher}. Molecular attributes such as local
structure and side chain conformations in these systems strongly
impact transport and device characteristics. In the PF family, poly
(9,9-(di \textit{n, n}-octyl) fluorene) (PF8) and poly (9,9-(di
ethyl hexyl) fluorene) (PF2/6) have received a lot of attention
mainly due to their mesomorphism and crystalline phases
\cite{chen,knaapila,tanto,arif_PRL}.

A controversial subject in PFs has been the physical mechanism
behind a broad green band in the photoluminescence (PL) and
electroluminescence (EL). There seems to be a general consensus that
its origin lies in keto defect (9-fluorenone) sites. Such defects
can be accidentally incorporated into the $\pi$-conjugated PF
backbone due to the presence of nonalkylated or monosubstituted
fluorene sites during synthesis or as a result of a photo-oxidative
degradation process \cite{List02}. However, the exact mechanism of
the green emission is heavily debated; it has been attributed to a
direct emission from flourenone defects \cite{zojer}, an
intramolecular charge transfer complex \cite{dias}, and due to
inter-chain excited species \cite{sims}.

\begin{figure}
\unitlength1cm
\begin{picture}(4.0,7)
\put(-2,-0.5){ \epsfig{file=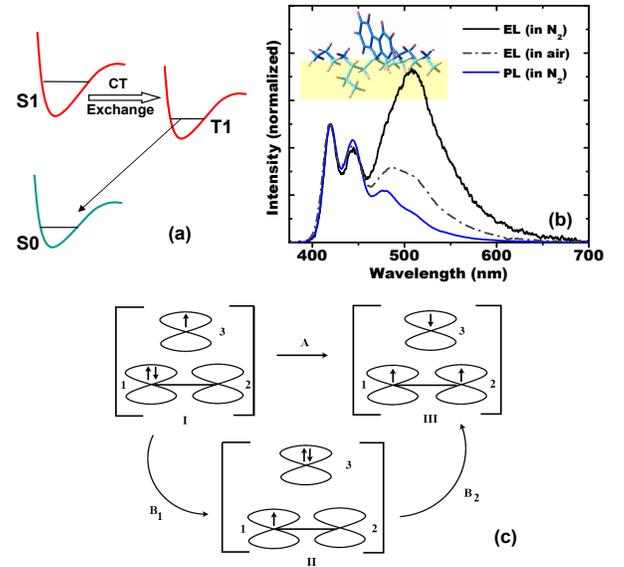, angle=0, width=8.5cm,
totalheight=8.5cm}}
\end{picture}
\caption{(a) Schematic of the conversion of singlet to triplet
states. (b) EL and PL spectra from a PF2/6 LED. The solid lines are
the EL and PL spectra from a device fabricated in N$_2$ atmosphere
and the dotted line represents the EL spectrum from an LED
fabricated under ambient conditions. Inset shows the structure of a
fluorene monomer with ethyl-hexyl side group. (c) The charge
transfer (B$_1$ and B$_2$) and the exchange (A) pathway for
conversion of the singlet excited state of an ethylene molecule
(1-2) to the triplet state by a paramagnetic impurity (3) are
depicted here. The charge transfer pathway is mediated by the
transfer integrals ($t_{31}$ for B$_1$ and $t_{23}$ for B$_2$) and
the exchange pathway is mediated through the two-electron integral
[$13|23$] in Eq.2.} \label{figure1}
\end{figure}

The inset of Fig. \ref{figure1}(b) shows a monomer unit of PF2/6;
the bridging carbon atom is the site of a keto defect where an
oxygen atom replaces the ethyl-hexyl group. The nature of the PL and
EL spectra in PFs due to these defects may also differ. Differences
in PL and EL are often observed in smaller molecules
\cite{kalinowski}. A recent work shows that charge-dipole
interactions stabilize energy states which are lower than the
singlet excited states, accounting for a red-shifted EL emission
\cite{sukrit}.

In this Letter we show that charge transfer (CT) and exchange
pathways cause mixing of the excited singlet ($S_1$) and the triplet
($T_1$) states, as shown schematically in Fig. \ref{figure1}(a). It
opens up possible decay channels from $T_1$ to the singlet ground
state ($S_0$), which is otherwise a forbidden transition. This
phenomenon is more generic than previously thought and may account
for the low energy emission mechanism in a host of short and
long-chain CP and molecules. The intensity of the low-energy EL
spectrum (490 nm-560 nm) in PF2/6 LEDs differ dramatically for
samples prepared in air (O$_2$), N$_2$, and doped with paramagnetic
impurities (PMs). A comparison of our experimental results with
theory clearly shows that the low energy EL emission can be
attributed to $T_1 \rightarrow S_0$ type transitions, and not
necessarily to keto defects as suggested in previous works.

The LEDs were fabricated by first spincoating 80 nm of
poly(ethylenedioxythiophene)-poly(styrenesulfonate) (PEDOT-PSS) onto
patterned indium tin oxide (ITO) glass slides, on top of which 100
nm of PF2/6 was spincoated from a toluene solution (10 mg/ml), and
capped by Ca/Al. Each sample supported 15 devices with area 4
mm$^2$, and was encapsulated prior to measuring EL and PL. The
devices under N$_2$ atmosphere were fabricated inside a N$_2$
glovebox including the cathode evaporation and encapsulation. The
ambient condition LEDs were obtained by first spincoating PF2/6 in
an ambient atmosphere, followed by the rest of the steps in a N$_2$
glovebox. Furthermore, metal doped LEDs were fabricated by
introducing trace quantities of Cu (II) (copper perchlorate) and a
Pd complex in the PF2/6 solution, both in ambient and N$_2$
atmosphere. Introduction of PMs in PF2/6 results in an electron-hole
recombination (e-hR) or  intersystem crossing (ISC) to the triplet
states, mediated by both exchange and CT mechanisms.

Our theoretical model system consists of a finite polyene chain and
a PM.  The e-hR process leading to excited singlet $S_n$ or triplet
state $T_m$ of a polyene is modeled with and without a paramagnetic
atom. The initial positively and negatively charged polyenic states
are the eigenstates within the Pariser-Parr-Pople (PPP) model
\cite{pople} with standard parameters and Ohno potentials
\cite{ohno} for long range Coulomb interactions. The metal atom is
described by a single orbital or two degenerate orbitals, with one
unpaired electron in both cases. The PPP Hamiltonian for the
molecule is given by \cite{soos}:
\begin{eqnarray} \label{1}
H^{\scriptscriptstyle PPP}_{polymer} & =&
\sum_i \alpha_i\hat{E}_{ii} + \sum_{\langle ij \rangle}t_{ij} \hat{E}_{ij}
+ \frac{1}{2}\sum_i U_i \hat{n}_i ( \hat{n}_i - 1)
\nonumber\\
&+& \sum_{i>j} V_{ij} (\hat{n}_i - z_i)(\hat{n}_j - z_j),
\end{eqnarray}
where
$\hat{E}_{ij}=\sum_{ij}\hat{a}^\dagger_{i,\sigma}\hat{a}_{j,\sigma}$;
$\hat{a}^\dagger_{i,\sigma}$, $\hat{a}_{j,\sigma}$, and $\hat{n}_i$
are the usual fermionic creation, annihilation, and number
operators, respectively. $\alpha{_i}$ and $t_{ij}$ are respectively
the site energies and hopping integrals, $\langle ij \rangle$
denotes the bonded pair, $U_i$ denotes on-site correlations, and
$V_{ij}$ the intersite interactions. The Hamiltonian of the magnetic
impurity is $H_{\rm
P}=\sum_{\sigma}\alpha_1\hat{a}^\dagger_{1,\sigma}\hat{a}_{1,\sigma}+(U_{\rm
P}/2) \hat{n}_{\rm P}(\hat{n}_{\rm P}-1)$. The total spin of the
full system (impurity and the polyene radical ions, $M^+$ and $M^-$)
are conserved, although the individual moieties can undergo a change
in spin state. The e-hR in the absence of PM occurs due to the CT
type interactions between the chains. In the presence of PM the
recombination occurs due to both one electron (CT pathways) and
two-electron (exchange pathways) interactions
\cite{tsubomura,hoijtink} between the impurity and the polymer
chain, schematically shown in Fig. 1(c). Both these pathways are
incorporated through the interaction Hamiltonian,
\begin{equation} \label{2}
H_{\rm int} = \sum_{ij} t'_{ij}\hat{E}_{ij} +
\frac{1}{2}\sum_{ijkl}[ij|kl] \left( \hat{E}_{ij}\hat{E}_{kl} -
\delta_{jk} \hat{E}_{il}\right),
\end{equation}
where $t'_{ij}$ is the one-electron transfer integral giving rise to
the CT process and $[ij|kl]$ is the two electron integral in charge
cloud notation.

Wave-packet propagation technique is employed to study the
electronic process involving the polyene ion-radicals and the
magnetic impurity. The initial state of the e-hR process is modeled
as a direct product of the ground states of a positive ($M^+$) and a
negative ($M^-$) polyene radical ion along with a PM, and is given
by
\begin{eqnarray} \label{3}
\Psi_{e-hR}(0)
=\frac{1}{\sqrt6}|\frac{1}{2},-\frac{1}{2}\rangle_{M^+}\otimes|\frac{1}{2},\frac{1}{2}\rangle_{M^-}\otimes|\frac{1}{2},\frac{1}{2}\rangle_{PM}
\nonumber\\
+\frac{1}{\sqrt6}|\frac{1}{2},\frac{1}{2}\rangle_{M^+}\otimes|\frac{1}{2},-\frac{1}{2}\rangle_{M^-}\otimes|\frac{1}{2},\frac{1}{2}\rangle_{PM}
\nonumber\\
-\frac{2}{\sqrt6}|\frac{1}{2},\frac{1}{2}\rangle_{M^+}\otimes|\frac{1}{2},\frac{1}{2}\rangle_{M^-}\otimes|\frac{1}{2},-\frac{1}{2}\rangle_{PM}
\end{eqnarray}
All states are labeled by the spin quantum numbers $S$ and $M_s$.
$|1/2,-1/2\rangle_M{\pm}$ is the ground state of the positive
(negative) ion-radical expressed as linear combination of the
valence bond (VB) functions which span the doublet space of the
ion-radical Hamiltonian. The wave-packet, $\Psi_{e-hR}(0)$, is time
evolved in discrete time steps by the time-dependent Schr\"{o}dinger
equation using the multistep differencing scheme (MSD) \cite{askar},
with the total Hamiltonian given by $H^{\scriptscriptstyle
PPP}_{M^+}+H^{\scriptscriptstyle PPP}_{M^-}H_{\rm P}+H_{\rm int}$.
The evolved state, $\Psi_{e-hR}(t)$, is then projected on to the
direct product of excited singlet (for singlet pathway) or triplet
(for triplet pathway) states, the singlet ground state of the
neutral molecule and the PM.

The probability for polaron recombination, $P_{T{_m}/S{_m}}$,
through triplet/singlet pathways is given by
$|\langle\Psi_{e-hR}(0)|\Psi^{\scriptscriptstyle
S/T}_{e-hR}(t)\rangle|^2$. Phosphorescence intensity is a measure of
the cross-section for radiative decay of the lowest triplet to the
ground state. The lowest triplet state is obtained from any of the
higher lying triplet states via internal conversion in the triplet
manifold according to Kasha's rule \cite{kasha}. The time integrated
yield of the triplet/singlet channel, $I_{T/S}\equiv
\sum_m\int_{0}^{T_{max}}P_{T{_m}/S{_m}}(t)dt$, is thus a measure of
the phosphorescence/fluorescence intensity in EL measurements.

\begin{figure*} \unitlength1cm
\includegraphics[angle=0, width=16cm]{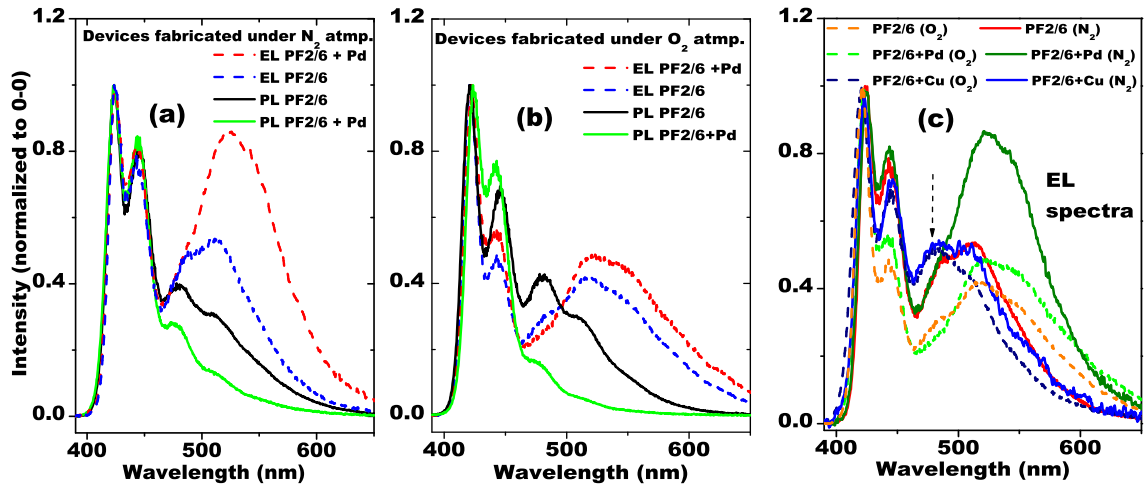}
\caption {EL and PL spectra of PF2/6 LEDs (a) fabricated in N$_2$
atmosphere, with and without Pd and (b) fabricated under ambient
conditions, with and without Pd. (c) EL spectra of all PF2/6 LEDs
fabricated in ambient and N$_2$ atmosphere; the dashed spectra
represent devices that were fabricated under ambient conditions. }
\label{figure2}
\end{figure*}

All EL and PL spectra have been normalized to the 0-0 vibronic peak.
Figure \ref{figure1}(b) compares the EL spectra of two PF2/6 LEDs,
one prepared under ambient condition and the other prepared in N$_2$
atmosphere. The intensity of green emission at 514 nm is
substantially higher for the device prepared in  N$_2$ atmosphere;
the green EL emission is quenched for devices prepared in O$_2$
atmosphere. Two batches of PF2/6 were used in this study. Figure
\ref{figure2} represents the EL and PL spectra from PF2/6 devices
(with PMs) utilizing the slightly higher mol. wt. sample compared to
Fig. \ref{figure1}.

 Figure \ref{figure2} (a) and (b) show the EL and PL spectra from Pd complex incorporated PF2/6
devices fabricated under N$_2$ and ambient atmosphere, respectively.
Fig. \ref{figure2} (c) shows all EL spectra including the devices
where trace concentration of Cu(II) was introduced. Although the
green EL emission from the high mol. wt. sample is lower than the
low mol. wt. sample (Fig. \ref{figure1} (b)) in N$_2$ atmosphere,
the devices prepared with the high mol. wt. sample under O$_2$
atmosphere have reduced green emission. The intensity of the green
emission is very much lower in the PL spectrum compared to the EL
spectrum in both cases. If the emission was due to oxidative
defects, one would expect the devices prepared in air to show a
higher  EL intensity. Besides, the PL and EL spectra would not be so
drastically different as observed.

The EL and PL spectra shown in Figs. \ref{figure1} and \ref{figure2}
are representative of at least 50 devices. We note that the PL
spectra are obtained from the actual devices and since the devices
were encapsulated, there is no additional occurrence of
photo-oxidation during the measurement. Incorporation of PMs quench
the PL emission in both N$_2$ and ambient environments (solid green
lines in Fig. \ref{figure2} (a) and (b)). A paramagnetic metal
significantly enhances the triplet exciton yield; hence it is not
surprising that the green emission is enhanced for Pd complex
incorporated LEDs fabricated under N$_2$ and ambient conditions
compared to as-is PF2/6 LEDs. The 514 nm emission is significantly
higher for the N$_2$ fabricated device compared to the device
fabricated under ambient atmosphere. This is expected since the
presence of oxygen is known to quench the triplet excitons
\cite{park}.

The green emission is very broad (490-560 nm) spanning the region of the 0-2 vibronic peak. The EL emission of the ambient PF2/6 device has a
clearly resolved 0-2 vibronic peak at 480 nm, shown by the dotted line with a shoulder at 514 nm; the latter is further enhanced in the device
fabricated in N$_2$ atmosphere (Fig. \ref{figure2} (c)). The EL spectra of the two Cu(II)-incorporated PF2/6 LEDs are shown in Fig.
\ref{figure2} (c). The 514 nm emission is slightly reduced for the device fabricated in O$_2$.

Our experimental results are corroborated by the theory very well.
The singlet and triplet state yields for butadiene calculated in
presence and absence of PM are presented in Table I. The singlet to
triplet state conversion by ISC pathway in presence of paramagnetic
metal is also performed for polyenes of different sizes; only the
octatetraene results are presented here. The polyene chains are
modeled using standard PPP parameters for Carbon \cite{soos}. The
butadiene units are placed end-on, 3.0{\AA} apart and the
paramagnetic site is located at 4.0{\AA} on the perpendicular
bisector of the full polyene system. The PM is modeled using either
one or two d-orbitals with the Hubbard potential,  U = 8.0 eV,
$\alpha$ varying between -5.0 eV to 5.0 eV, bond-bond repulsion
integrals $[ij|kl]$ $\sim$ 0.5 eV for the nearest neighbor pair
$\langle i j\rangle$, $\langle k l\rangle$, and $t'$ $\sim$ 0.2 eV.
These parameters are consistent with photo-emission spectroscopy
data \cite {smith} for 3d/4d orbitals of metals which are 3 to 5 eV
above the carbon 2p orbitals. The triplet yields are only weakly
dependent upon the site energies. The higher yield of triplets has
been observed when unpaired electron on the paramagnetic moiety is
in a pair of degenerate orbitals. In the absence of PM the e-hR
process yields more singlets than triplets. The triplet yield in
presence of PM is much higher in e-hR process compared to the ISC
pathway. Furthermore, the singlets formed in EL can be converted
into triplets by ISC mechanism in the presence of PMs, which
subsequently undergo internal conversion to the low energy $T_1$
from which phosphorescence may occur.

Our experimental results are consistent with the above; the
intensity of the 514 nm EL peak is higher when doped with Pd
complexes compared to Cu(II) salts (Fig. 2 c). In PL, the ISC is the
only mechanism for the production of triplets. For this the singlet
exciton must be in the vicinity of the PM, which at low impurity
concentration is less probable. Hence we do not observe enhancement
in PL green emission intensity. We have also calculated the
quenching probability of triplet states in the presence of a
paramagnetic oxygen molecule. Our calculations show that there is
substantial quenching of the triplet excitons by O$_2$ which we
attribute to the observed low intensity of green emission in samples
prepared under ambient conditions with or without metal centers.

\begin{table}
\caption{Singlet and triplet yields (I$_{T/S}$) in e-hR (for a pair
of butadienes) and ISC (for octatetraene) in presence$^1$ and
absence$^2$ of PM. Triplet yields in ISC are from the first excited
state (2A$_g$).} \label{table1}
\begin{ruledtabular}
\begin{tabular}{cccccc}
Pathways & Site energy & \multicolumn{2}{c}{
   I$_T$} & \multicolumn{2}{c}{I$_S$} \\
& of PM(eV) & Nondeg. & Deg. & Nondeg. & Deg.  \\ \hline
ISC & 5.0  & 0.29 & 0.23& &\\
e-hR$^1$ & 5.0    &  12.43 & 0.89 & 4.91 & 0.83 \\
e-hR$^2$ &- & \multicolumn{2}{c}{0.44}  & \multicolumn{2}{c}{1.82}\\
\end{tabular}
\end{ruledtabular}
\end{table}

Geminate recombination of polaronic pairs result in the formation of
both singlets and triplets, in systems with or without defects. Our
calculations show that oxygen molecules annihilate the triplets
while PMs generate triplets with high probability either from
singlet excited states via ISC or from e-hR process. The broad green
emission may be attributed to vibronic broadening and the triplet
states of keto defects. ZINDO (Zerner's Intermediate Neglect of
Differential Overlap) calculations carried out for a fluorene dimer
with one or no keto defect show that the system with keto defect has
a triplet state at 2.37 eV (close to our experimental results) while
that without a keto defect has a triplet state at 2.89 eV. Thus the
green emission could occur from the triplet state created in the
vicinity of a keto defect due to migration of the free triplet to a
keto defect site. Electrophosphorescence has indeed been observed at
low temperatures in PFs using time-resolved detection techniques
\cite{sinha}; our work clearly shows a large cross-section of
triplets in the eh-R process, particularly in the presence of PM.

In summary, our studies show that the low energy EL emission in the
PF system is from triplet excitons. This is corroborated by
experimental studies on LEDs prepared in N$_2$ atmosphere and
ambient conditions with different transition metal and ion dopants,
and theoretical studies of ISC and electron hole recombination in
the presence of PMs as well as triplet state quenching by O$_2$. The
dopants provide both exchange and charge transfer pathways to
convert singlet to triplet excitons in PF systems with and without
keto defects. Such processes are not just restricted to PF systems
but may explain the low energy EL emission in a number of conjugated
polymers.
%**********************************************************************************

\begin{acknowledgments}
We thank Ulli Scherf for providing the PF2/6 sample and Satish Patil
for valuable discussions. We gratefully acknowledge the NSF for
support through grant Nos. ECCS-0523656 and 0823563. Work in India
was supported by DST, India through grant no SR/S2/CMP-24/2003.

\end{acknowledgments}

%***************************************************************************
%***************************************************************************

%***************************************************************************
%***************************************************************************
%***************************************************************************


\begin{thebibliography}{100}
\bibitem{neher}
D. Neher Macromol. Rapid Commun. {\bf 22}, 1365 (2001).

\bibitem{chen}
S. H. Chen, A. C. Su, C. H. Su, and S. A. Chen, Macromolecules {\bf38}, 379 (2005).

\bibitem{knaapila}
M. Knaapila, R. Stepanyan, B. P. Lyons, M. Torkkeli, T.P.A. Hase, R. Serimaa, R. G\"{u}ntner, O.H. Seeck U. Scherf, and A. P. Monkman,
Macromolecules {\bf38} 2744 (2005).

\bibitem{tanto}
B. Tanto, S. Guha, C.M. Martin, U. Scherf, and M.J. Winokur, Macromolecules {\bf37}, 9438 (2004).

\bibitem{arif_PRL}
M. Arif, C. Volz, and S. Guha, Phys. Rev. Lett. {\bf96}, 025503 (2006).

\bibitem{List02}
E.J.W. List, R. Guentner, P.S. de Freitas, and U. Scherf, Adv. Mater. {\bf14}, 374 (2002).

\bibitem{zojer}
E. Zojer, A. Pogantsch, E. Hennebicq, D. Beljonne, J. L. Brédas, J.
Chem. Phys. {\bf117}, 6794 (2002).

\bibitem{dias}
F.B. Dias, M. Maiti, S.I. Hintschich, and A.P. Monkman, J. Chem.
Phys. {\bf122} 054904 (2005).

\bibitem{sims} M. Sims, D.D.C. Bradley, M. Ariu, M. Koeberg, A.
Asimakis, M. Grell, and D.G. Lidzey, Adv. Funct. Mater. {\bf14}, 765
(2004)

\bibitem{kalinowski}
J. Kalinowski, G. Giro, M. Cocchi, V. Fattori, P. DiMarco, Appl. Phys. Lett. {\bf76}, 2352 (2000).

\bibitem{sukrit}
Z.G. Soos, S. Mukhopadhyay, and S. Ramasesha, Chem. Phys. Lett. {\bf442}, 285 (2007).

\bibitem{pople}
J. A. Pople, Trans. Faraday Soc. {\bf49}, 1375 (1953); R. Pariser, R. G. Parr, J. Chem. Phys. {\bf21}, 767 (1953).

\bibitem{ohno}
K. Ohno, Theor. Chim. Acta. {\bf2}, 219 (1964).

\bibitem{soos}
Z.G. Soos and S. Ramasesha, Phys. Rev. B {\bf29}, 5410 (1984).

\bibitem{tsubomura}
H. Tsubomura, R. S. Mülliken, J. Am. Chem. Soc. {\bf82}, 5966 (1960).

\bibitem{hoijtink}
G.J. Hoijtink, Mol. Phys. {\bf3}, 67 (1960).

\bibitem{askar}
A. Askar and A.S. Cakmak, J. Chem. Phys. {\bf68}, 2794 (1978).

\bibitem{kasha}
M. Kasha, Disc. Faraday Soc. {\bf9}, 14 (1950).

\bibitem{park}
J.H. Park, Y.T. Lim, O.O. Park, J.K. Kim, J-W. Yu, and Y. C. Kim, Chem. mater. {\bf16}, 688 (2004).

\bibitem{smith}
N. V. Smith, G. K. Wertheim, S. Hüfner, M. M. Traum, Phys. Rev. B {\bf10}, 3197 (1974).

\bibitem{sinha}
S. Sinha, C. Rothe, R. G\"{u}ntner, U. Scherf, and A.P. Monkman,
Phys. Rev. Lett. {\bf90}, 127402 (2003).


\end{thebibliography}
\end{document}